\documentclass[10pt,preprint,twocolumn]{aastex}
\usepackage{graphicx}% Include figure files
\usepackage{dcolumn}% Align table columns on decimal point
\usepackage{bm}% bold math
\usepackage{multicol}
\usepackage{flafter}
\newcommand{\beq}{\begin{equation}}
\newcommand{\eeq}{\end{equation}}
\newcommand{\bea}{\begin{array}}
\newcommand{\eea}{\end{array}}

\begin{document}

\title{Dynamics and Eccentricity Formation of Planets in OGLE-06-109L System}
\author{Su Wang, Gang Zhao and Ji-Lin Zhou$^*$}
\affil{Department of Astronomy, Nanjing University, Nanjing 210093, China \\
($^*$zhoujl@nju.edu.cn)}
% \twocolumn[
%\begin{@twocolumnfalse}
%\maketitle
%\documentstyle[twocolumn]{article}

\begin{abstract}
Recent observation  of microlensing technique reveals  two giant
planets at  2.3 AU and 4.6 AU around the star OGLE-06-109L. The
eccentricity of the outer planet ($e_c$) is estimated to be
0.11$_{-0.04}^{+0.17}$, comparable to  that of Saturn (0.01-0.09).
%with unknown eccentricity for the inner planet .
The similarities between the OGLE-06-109L system and the solar
system indicate that they may have passed through  similar
histories during their formation stage.  In this paper we
investigate the dynamics and formation of the orbital architecture
in the OGLE-06-109L system. For the present two planets with their
nominal locations, the secular motions are stable as long as their
eccentricities ($e_b,~e_c$) fulfill $e_b^2+e_c^2 \le 0.3^2$.
Earth-size bodies might be formed and are stable in the habitable
zone (0.25AU-0.36AU) of the system. Three possible scenarios may
be accounted for formation
 of $e_b$ and $e_c$: (i) convergent migration of two planets and the 3:1 MMR trapping;
 (ii) planetary scattering; (iii) divergent migration and the 3:1 MMR crossing.
As we showed that the probability  for the two giant planets in
3:1 MMR is low ($\sim 3\%$),  scenario (i) is less likely.
According to models (ii) and (iii), the final eccentricity of
inner planet ($e_b$) may oscillate between [0-0.06], comparable to
that  of Jupiter (0.03-0.06). An inspection of $e_b$, $e_c$'s
secular motion may be helpful to understand which model  is really
responsible for the eccentricity formation.

\end{abstract}

%\end{@twocolumnfalse}
%  ]
%

 \keywords{(\textit{stars:}) planetary systems: formation -- solar system: formation -- stars:
individual (OGLE-06-109L) }

%\begin{multicols}{2}
\section{Introduction}

To date, more than 370 exoplanets are detected mainly by Doppler
radial velocity measurements(\footnote{http://exoplanet.eu/}),
 among them  9 planets are revealed
by gravitational microlensing (Udalski et al. 2005; Beaulieu et
al. 2006; Bennett et al. 2006, 2008; Gould et al. 2006; Gaudi et
al. 2008; Dong et al. 2009; Janczak et al. 2009). The use of
microlensing   technique to  planet-searching is based on the idea
of general relativity  that light passing through a mass should
bend as if it passes a lens (Mao \& Paczynski 1991). The major
advantage of the technique is that it favors to detect planets in
moderate distance to the host star ($\sim$ 3 AU), which is
complimentary to the radial velocity measurements. Among the
planets detected by microlensing, OGLE-06-109L system is the first
one with observed multiple planets.  It is 1490 pc away from the
Sun, with a star of $\sim 0.5~ M_\odot$ (solar mass) and  two
planets of $0.71~M_J$ (Jupiter mass) and 0.27 $M_J$ in the orbits
of 2.3 AU and 4.6 AU, respectively (Gaudi et al. 2008).  Table 1
lists their nominal orbital elements.

Several features of the system make it an analogy of the solar
system.  (i)  The two planets have positions and masses similar to
those of Jupiter and Saturn up to  scale changes. (ii) The fitted
eccentricity of  OGLE-06-109L c  is modest
(0.11$_{-0.04}^{+0.17}$), comparable with those of Jupiter
(0.03-0.06) and Saturn (0.01-0.09) during their secular evolution,
although the eccentricity of  OGLE-06-109L b  is unknown. (iii)
While Jupiter and Saturn's orbits are closing to $5:2$ mean motion
resonance (MMR), OGLE-06-109L b and c may be close to $3:1$ MMR.
(iv) All these four giant planets are  located well outside the
snow lines of their systems ($\sim 0.68$ AU for OGLE-06-109L and
$\sim 2.7$ AU for  solar system),   indicating that, unlike the
most observed extra-solar systems with hot Jupiters, the migration
of these four giant planets were not so efficient. (v) The
habitable zone of the OGLE-06-109L system,  [0.25AU-0.36AU], may
have stable orbits.

Considering most of the observed exoplanets
have close-in orbits with an average eccentricity $\sim 0.2$ (Udry
\& Santos  2007),  the  investigation of formation scenario in  OGLE-06-109L
system may bridge the gap between the solar system and the most observed  exoplanet systems.
  Another aim of the paper is to predict the eccentricity of OGLE-06-109L b.
  Due to the significant uncertainties  in  orbital determination
of microlensing technique ($\sim 10\%$,  S.  Mao, private
communication), the orbital parameters of the two detected planets
 in OGLE-06-109L system  are poorly known. A precise
determination  by other means like radial velocity is still
impossible due to the great distance ($\sim 1490$ pc) between
OGLE-06-109L and the Sun.
 So the investigation of their
formation and dynamics is  helpful to reveal their orbital
parameters, especially their eccentricities.

The mechanism of eccentricity excitation is not fully understood,
especially in single planet systems. During the early stage of
planet formation, disk-planet interaction tends  to damp the
eccentricity of the planet. According to the  linear theory, the
planet exerts torques on a dynamically cold disk ($c\ll r\Omega $,
where $c,~r,~\Omega$ are the sound speed of gas, the orbital
radius and angular velocity of planet motion, respectively) mainly
at the Lindblad (LR) and corotation (CR) resonances (Goldreich \&
Tremaine, 1979, 1980; Ward 1988). Only non co-orbital LRs
 can excite planetary eccentricity, while
CRs and co-orbital LRs  damp the eccentricity. In the linear
regime, torques from the latter dominate the evolution, so the
planetary eccentricity  is damped, unless  the
 planet is massive enough to clear the local gas disk (Lin \& Papaloizou 1993).
Two-dimensional hydrodynamical simulations show that the critical
mass ($M_{\rm crit}$) of the planet above which its eccentricity will be
excited under disk tide  is
 $\sim 20~ M_J $, and $M_{\rm crit}$ might be
reduced into the range of the  observed extrasolar planets at a
very low disc viscosity (Papaloizou et al. 2001). During the later
stage of planet formation, the depletion of the gas disk will
increase the eccentricity of  a planet through the sweeping of secular
resonance (Nagasawa et al. 2003).

For multiple planetary systems, there are mainly several scenarios
that will excite the eccentricities of the planets:

(i) Convergent migration and resonance trap between two planets. A
planet in a gaseous disk will migrate inward either due to the
imbalance of LR torques  or co-evolute with the viscous disk when it
is massive enough to open a gap around it. In the case that the
migration speed of inner planet is slower than that of the outer
one, or  the inner one is stalled due to the clear of nearby gas, a
trap into MMR between the two planets is possible, which may result
in the increase of both eccentricities (Lee \& Peale 2002; Kley
2003).  Due to the trap of MMR,   their eccentricities remain
oscillating around  moderate values ($\sim 0.1-0.3 $) at the end of
evolution.
 The configurations of  GJ 876 b-c in 2:1 MMR and  55 Cnc  b-c in 3:1 MMR are believed
to be formed in this way.

(ii) Planetary scattering. During the formation stage of planets,
protoplanetary cores may undergo close encounters with the
planets, causing ejections of the cores  and the eccentricities
excitation for the survival planets. Secular interactions between
the survival planets ($m_b$ and $m_c$) may result in oscillations
of eccentricities between 0 and a finite value ($\sim 0.1$). Such
a motion is near the separatrix of libration and circulation of
difference of perihelion longitude ($\Delta \varpi_{bc}$) in the
eccentricity plane ($e_b e_c \cos \Delta \varpi_{bc}$, $e_be_c
\sin \Delta\varpi_{bc}$),  so it is called a near-separatrix
motion (Barnes \& Greenberg 2006, 2008).
 This  model can account for  the eccentricity properties of  the $\upsilon$ Andromedae
system (Ford et al. 2005).

(iii) Divergent migration and MMR crossing under interaction with
planetesimal disk. After circumstellar disk depletes due to disk
accretion, photoevaporation or planet formation within $\sim 3$
Myrs (Haisch et al. 2001), planets may undergo migration through
angular momentum exchanges with residue embryos and planetesimals
(Fernandez \& Ip 1984; Malhotra 1993; Hahn \& Malhotra 1999). In
the solar system, numerical simulations show that, Jupiter will
drift inward, while Saturn, Uranus, Neptune may migrate outward,
resulting in a divergent migration. During the migration, the
cross of 2:1 MMR between Jupiter and Saturn
  excites the eccentricities of four giant planets (Tsiganis et al. 2005), which may result in the formation
of Trojans population of Jupiter and Neptune (Morbidelli et al.
2005), the later heavy bombardment of the terrestrial planets
(Gomes et al. 2005), and the architecture of Kuiper belt (Levison
et al. 2008).

(iv) Slow diffusion due to planetary secular perturbation. During
the final stage of planet evolution when planets are almost formed
in well separated orbits, secular perturbations between them
result in a slow increase of stochasticity of the system. This
procedure can be approximated as a random walk in the space of
velocity dispersion, and the  resulting eccentricities of the
planets obey a Rayleigh distribution, which agrees with the
statistics of  the eccentricities for the observed exoplanets
(Zhou et al 2007).
 The difference between this and the previous
planetary scattering scenario is that,  slow diffusion model  may
occur in a much longer time span, it is effective especially in
the later stage of planet evolution when the planetary orbits are
well separated and there is no  violent scattering events.

 In this paper, we investigate  the
eccentricity formation scenarios and dynamics of the OGLE-06-109L
system through N-body simulations, with focuses on the following
topics: (a) the origin of the eccentricities for  the two giant
planets,  (b)  the stability of the present configuration, (c) the
possible existence of planets in  the habitable zone and the outer
region. The  eccentricity formation scenarios revealed in this
paper  can  be extended to other multiple planetary systems. The
paper   is organized as follows. In section 2, the dynamics of the
present system is investigated, with much attention paid on the
stability of the two giant planets system, the inner and outer
regions. Then in section 3, we study the various scenarios (i, ii,
iii) as mentioned above to account for the eccentricity formation
of the two giant planets. Conclusions and discussions presented in
the final section.
 \vspace{-1cm}
\section{Dynamics and Stability of Nominal System}
Hereafter we denote $m_b$ and $m_c$ as OGLE-06-109L b and
OGLE-06-109L c, and the corresponding orbital elements are
detached by a subscript  $b$ or $c$, respectively.
    As the nominal orbital periods of $m_b$, $m_c$ are close to $3:1$ MMR (with period ratio
$2.8\pm0.7$, Table 1), we first check the possibility of the
present two planets in 3:1 MMR within the observational error.
Then the dynamics and stability of the OGLE-06-109L system are investigated.

\begin{table*}[htp]
\centering
 \caption{\footnotesize{Orbital parameters of OGLE-06-109L planetary
system.}}\label{tbl-1}
\footnotesize{
\begin{tabular}{cccccc}
\tableline\tableline Planet & $a$ & $P$ & $e$ & $\varpi$ & $m$\\
&(AU)&(days)&&&$(M_J)$
\\
\tableline
b &$2.3 ~(\pm 0.2)$ &$1825 ~(\pm 365)$ &-  &- &$0.71 ~(\pm0.08)$ \\
c &$4.6 ~(\pm 0.5)$ &$5100 ~(\pm 730)$ &$0.11
~(_{-0.04}{^{+0.17}})$ & - &$0.27 ~(\pm0.03)$ \\ \tableline
\end{tabular}}
%% Any table notes must follow the \end{tabular} command.
\tablenotetext{}{from Gaudi et al. (2008) and
http://exoplanet.eu.}

\end{table*}
 \vspace{-0.5cm}
\subsection{3:1 mean motion resonance?}
 We assume that the masses and orbital elements of the two
planets obey normal distributions,
\begin{equation}
P(x)=\frac{1}{\sigma\sqrt{2\pi}}\exp\left[-\frac{(x-x_0)^2}{2\sigma^2}\right],
\label{px}
\end{equation}
with $x_0$ the nominal value of $x$, and $\sigma$  the
observational error. The distribution of eccentricity is obtained
by $e=\sqrt{x_1^2+x_2^2}$, where  $x_1,~x_2$ are two independent
Gaussian (\ref{px}) with $x_0=0$ and different dispersions
$\sigma_1,~\sigma_2$, and it is
 the standard Rayleigh distribution (Zhou et al. 2007)
when  $\sigma_1=\sigma_2$. We choose values of $\sigma_1$ and
$\sigma_2$ so that the most probable eccentricity of $m_c$ is
0.11, with  $1\sigma$ confidence interval $(0.07,~0.28)$. Also we
assume  $e_b$ follows a Rayleigh distribution with the most
probable value $({\bar e}_b)$. The distributions of the
inclinations  are the same with those  of  $e/2$. The remaining
angles  are randomly generated.

We carry out 15000 runs of simulations by integrating the full
motion of three bodies ($m_*$, $m_b$, $m_c$) in the
three-dimensional physical space up to
0.1 Myrs. %These simulations are divided into 3 groups according to
%the most probable values of $m_b$'s eccentricity (${\bar e}_b=0.04,~0.1,~0.2$).
  By checking whether any one of six resonant
angles ($3\lambda_c-\lambda_b-i\varpi_b-j\varpi_c-k\Omega_b-l\Omega_c$
 with $i,~j,~k,~l$ non negative integers and $i+j+k+l = 2$) librates,
we find  the probabilities that $m_b$, $m_c$ in 3:1 MMR are
$0.82\%, ~2.52\%,~ 1.88\%$ for ${\bar e}_b=0.04,~ 0.1,~ 0.2$,
respectively.  Thus the probability for  $m_b$ and $m_c$ in the
3:1 MMR is small within observational errors.

\subsection{Secular dynamics of two planets}

\begin{figure}[htp]
\vspace{-2cm}
%\begin{center}

\includegraphics[scale=0.7]{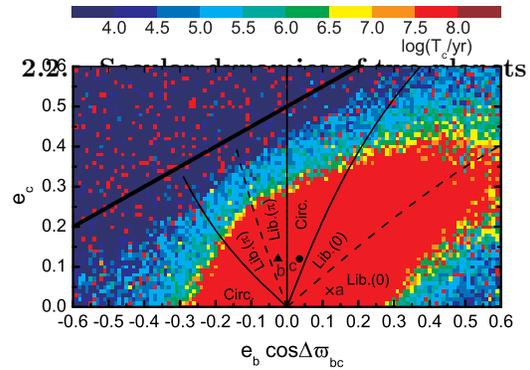}
\centering
 \vspace{-0.5cm}
% \epsscale{0.5}   \plotone{fg1.eps}
% \vspace{-5cm}
 \caption{{\footnotesize Various motions and orbital crossing timescale $T_c$
 for orbits initiated from the
 representative plane $(e_{b}\cos\Delta\varpi_{bc},~e_{c})$.
 In the right (left) half-plane, the initial $\Delta\varpi_{bc}=0$ ($\Delta\varpi_{bc}=\pi$, respectively).
  The dashed curves locate the equilibriums of the secular system with
$\Delta\varpi_{bc}(t) \equiv 0$ or $\Delta \varpi_{bc} (t)  \equiv
\pi$.
 The thin solid lines separate the
 domains that $\Delta\varpi_{bc}(t)$ librates about $0$/$\pi$ from
 circulation. The thick solid line in the upper-left corner
shows the boundary above which $a_b(1+e_b)< a_c(1-e_c)$. The
region in
 red is stable (without orbital crossing) in full three-body integration up to $10^8$
 years. The regions in blue  are unstable with $T_c\le 10^4$ years.
 Between them are  chaotic
 regions with  $ T_c\sim 10^{4-8}$ years.
Motions originating from points with notation (a-c) are plotted in
Fig. 2. }}
%\end{center}
\end{figure}
 To investigate the secular dynamics of the two planets,  we
adopt the method of representative plane of initial conditions
(Michtchenko \& Malhotra 2004).  Due to the existence of four
center of mass integrals, the planar three-body (the star and two
planets) system  is a Hamiltonian one with four degrees of
freedom, with the Hamiltonian function (e.g., Laskar \& Robutel 1995):
\begin{equation}
H=-\sum_{i=b}^c\frac{\mu_i^2m'^3_i}{2L_i^2}-G\frac{m_b m_c}{|{\bf
r}_b-{\bf r}_c|} + \frac{m_bm_c}{m_\ast}(\dot x_b\dot x_c+\dot
y_b\dot y_c),
\end{equation}
where $\mu_i=G(m_\ast+m_i)$, $m'_i=m_im_\ast/(m_\ast+m_i)$,   $
L_i =m'_i\sqrt{\mu_ia_i}$ ,  with $a_i$, $e_i$, ${\bf r}_i$  being
the semi-major axis, eccentricity, relative position vector  of
$m_b$  or $m_c$,  respectively. An averaged system is obtained by
 eliminating the short
periodic terms,
\begin{equation}
H_{\rm{sec}}=-\frac{1}{(2\pi)^2}\int_0^{2\pi}\int_0^{2\pi}
Hd\lambda_b\lambda_c, \label{Hsec}
\end{equation}
where $\lambda_b,~ \lambda_c$ are the longitudes of mean motion of
$m_b$ and $m_c$ respectively.
 Due to the D'Alembert's rule, only
$\Delta\varpi_{bc}=\varpi_c-\varpi_b$ appears in the average
Hamiltonian,  thus the secular system is integrable with one
degree of freedom in $\Delta\varpi_{bc}$ and its conjugate
momentum $K_b= L_b\left(1-\sqrt{1-e_b^2}\right)$ , and $\Delta
\varpi_{bc}$ either circulates or liberates around $0$ or $\pi$.
So  the information of secular motion with different initial
conditions can be presented in the representative plane
$(e_b\cos\Delta\varpi_{bc},~e_c)$, where $\Delta\varpi_{bc}$ is
fixed at either $0$ or $\pi$.

 Fig.1 shows the various motions in  the representative plane of
the OGLE-06-109L system, with the semi-major axes of the planets
being  the nominal values (Table 1). Two families of the
equilibriums  of the secular system (\ref{Hsec}),
 with $\Delta\varpi_{bc} (t) \equiv 0$ or $\pi $ are plotted as dashed lines.
 Nearby orbits  are with $\Delta\varpi_{bc} (t)$ librating around $0$ or $\pi$.
 Between the two libration regions are orbits with $\Delta\varpi_{bc}$-circulating,
 bounded by thin solid lines.
We also plot the orbital crossing time ($T_c$) of orbits
originating from the representative plane by full three-body
integrations (with other angles randomly chosen). $T_c$ is defined
as the minimum time that either $a_b>a_c$ occurs or the separation
of $m_b$ and $m_c$ is smaller than one mutual Hill radii.
 From Fig.1, we  see that apsidal alignment ($\Delta\varpi_{bc}  \approx 0$)  between
$m_b,~m_c$ can stabilize the interacting planets.

\begin{figure}[htp]
%\begin{center}
\centering
 \vspace{-0.5cm}
\includegraphics[scale=0.7]{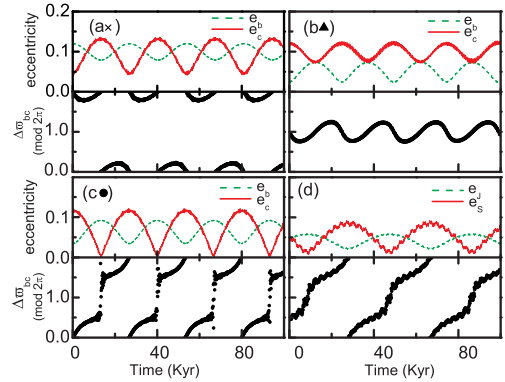}
 \vspace{-0.5cm}
 %\vspace{-11cm}
\caption{\footnotesize {Some typical evolution of orbits
originating from the representative plane of Fig.1. Panel (a)
shows the motion with $\Delta\varpi_{bc}$ librating around 0.
 Panel (b) is the motion of which $\Delta\varpi_{bc}$ librates around $\pi$. Panel
 (c) is  an example of near-separatrix motion:  $e_b $
  is periodically back to $\sim 0$ ,  which is similar to
the orbits of the Jupiter-Saturn system in Panel (d).}}
%\end{center}
\end{figure}

 Some typical motions originating from Fig.1
 are plotted in Fig.2.  Orbits (a), (b) are from the libration region
so that $\Delta \varpi_{bc}$ librates around $0$ or $\pi$,
respectively. Orbits  (c) are with $\Delta \varpi_{bc}$
circulating and $e_c$  is in a near-separatrix motion, i.e.,
oscillates between 0 and a finite value ($\sim 0.1$) periodically,
which is very similar to the  Jupiter-Saturn system in panel (d).

\subsection{Stable regions of the two-planet system}
\begin{figure}[htp]
%\begin{center}
\centering
%\hspace{2cm}
\includegraphics[scale=0.8]{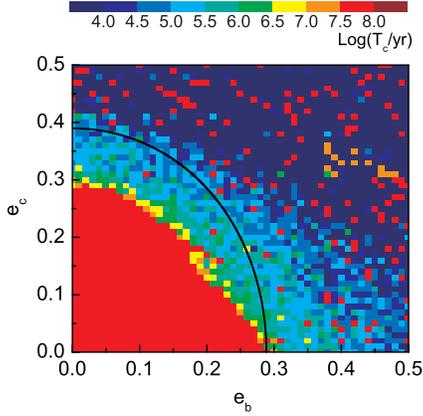}
 \vspace{-0.5cm}
 %\vspace{-11cm}
\caption{\footnotesize {Orbital crossing time $T_c$ in $e_b,~e_c$
plane for the two giant planets in the nominal locations (Table
1). All the other angles $\varpi_b,~\varpi_c,~M_b,~M_c$ are random
chosen. The solid curve plots Hill stability criterion given by
Eq.(\ref{Hill}).}}
%\end{center}
\end{figure}
We integrate the three-body system with a second-order WHM code
 (Wisdom \& Holman 1991) from the SWIFT
package (Levison \& Duncan 1994). The time step is set as the $1/12$ of
 the  period of the innermost orbit.
Taking  initial $a_b$ and $a_c$  with present nominal values,
we carry out 2500 runs of integrations with different initial eccentricities.
 All the angles including
$\varpi_b$ and $\varpi_c$ are random chosen. The orbital crossing
time $T_c $, defined as the minimum time that $a_b>a_c$ or
distance of $m_b$ and $m_c$ being smaller than one mutual Hill
radius, is presented in Fig.3. According to the results, the
two-planet ($m_b$ and $m_c$) system is stable ($T_c>10^8$yrs) only
if
\begin{equation}
e_b^2+e_c^2<0.3^2
\end{equation}
holds approximately. Let us compare this to another commonly used
stability.  Hill stability requires the ordering of the two
planets remain unchanging for all the time, which allows the outer
planet escaping to infinity. Topological studies of three-body
systems give the following sufficient criterion for
 two planets  being Hill stable (Marchal \& Bozis 1982; Gladman 1993),
  \beq
-\left(\frac{2m_{\rm{tot}}}{G^2m^3_{\rm{pair}}}\right)L^2E>
1+\frac{3^{4/3}m_bm_c}{m_\ast^{2/3}(m_b+m_c)^{4/3}}+\cdots,
\label{Hill} \eeq where $m_{\rm{tot}}=m_\ast+m_b+m_c$,
$m_{\rm{pair}}=m_\ast m_b+m_\ast m_c+m_b m_c$, $L$ and $E$ are the
total angular momentum and energy of the three-body system,
respectively. This criterion is also plotted in Fig.3. Note that
the Hill stability criterion gives a larger region as that of
$T_c>10^8$ years.  This implies that, the stability defined by
orbital crossing time $T_c> 10^8$ years is stronger than the Hill
stability, in the sense that,  our stability in terms of orbital
crossing
 requires the mutual distance of two planets
 being more than one mutual Hill radius all the time.

\subsection{Stability of fictitious planets}

We numerically integrate the orbits of a few hundred test
particles in the planetary system.  Such simulations enable us to
identify regions where low-mass companions can have stable orbits
(Rivera \& Haghighipour 2007; Haghighipour 2008). The test
particles are initially located in circular orbits coplanar with
two planets, with their mean longitudes randomly set. The orbital
evolution time is set as 100 Myrs. A particle is removed from the
simulation when its stellar  distance exceeds 30 AU, or it enters
the Hill sphere of either planet.
 We study two cases
 that either $m_b$ and $m_c$ are in 3:1 MMR or not.
 In the non-resonance cases, $m_b$ and $m_c$
 are initially located at the nominal elements (Table 1) with $e_b=0.06$ (a value
inferred from formation scenario in next section). To show the
dependence of stability on $\Delta \varpi_{bc}$, we let initially
$\varpi_b=0$, $M_b=0$, and $\varpi_c=0,~\pi/2,~\pi$. $M_c$ is
randomly set. In the 3:1 MMR cases,  $m_b,~m_c$  are set with
nominal parameters except $a_b=2.11445$ AU, $e_b=0.06$,
$\varpi_b=\varpi_c=M_b=0$, $M_c=\pi$, so that $m_b$ and $m_c$ are
initially in  3:1 MMR, and the three corresponding resonance
angles, $3\lambda_c-\lambda_b-2\varpi_b$,
$3\lambda_c-\lambda_b-\varpi_b-\varpi_c$,
$3\lambda_c-\lambda_b-2\varpi_c$, liberate around $\pi, ~0,~\pi$,
respectively.
\begin{figure}[htp]
%\begin{center}
\centering
 \vspace{-0.5cm}
\includegraphics[scale=0.5,width=3.5in]{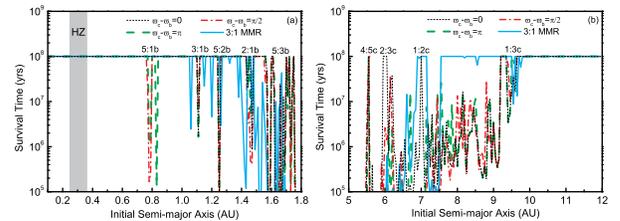}
 \vspace{-1cm}
 %\vspace{-11cm}
\caption{\footnotesize {Lifetime of test particles in OGLE-06-109L
system at different semi-major axes. Mean motion resonances are
marked on the top (e.g. 5:1b means the 5:1 MMR with $m_b$). The
grey band in (a) shows the extension of the habitable zone.}}
%\end{center}
\end{figure}

\begin{figure}[htp]
%\begin{center}
\centering
 \vspace{-0.5cm}
\includegraphics[scale=0.7]{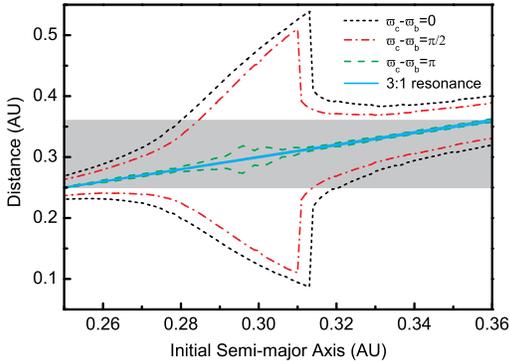}

\caption{\footnotesize{The largest and smallest stellar distances
of an Earth-mass planet  with different initial semi-major axes.
$m_b$ and $m_c$ are initially put  at the nominal locations and
eccentricities ($e_b=0.06$) except the 3:1 MMR case, where
$a_b=2.11445$ AU. The grey band is the extension of the habitable
zone.}}
%\end{center}
\end{figure}
{\em Inner region.} Fig.4a shows the survival time  of test
particles in inner system with different  initial semi-major axes.
When $m_b$ and $m_c$ are not in 3:1 MMR, test particles with initial $a \leq 1.5$ AU are
  stable in the sense of $T_c>100$ Myrs, except at 5:1, 3:1, 5:2, 2:1 MMRs with
$m_b$. The locations of these MMRs depend on $\Delta \varpi_{bc}$.
However, if $m_b$ and $m_c$ are in 3:1 MMR, the stable region is
reduce a bit to $a \leq 1.4$ AU.

{\em Outer region.} Fig.4b shows the survival time of test
particles in the outer system. When $m_b$ and $m_c$  are not in 3:1
MMR, the test particles are stable as long as $a \geq 9.7$ AU. However,
If the two planets are  in  3:1 MMR ,  the stable region is
enlarged to $a \geq 7.5$ AU except the 1:3 MMR with $m_c$.
It is interesting to note that, the 3:1 MMR between $m_b$ and $m_c$ increases
the stable region in the outer regions.

{\em Habitable zone.} Kasting et al. (1993) estimated the width of
the habitable zone (HZ), where an
Earth-like planet can have liquid water on its surface, around main sequence stars.
 For the
star with mass 0.5 $M_\odot$, the most conservative position of HZ
is 0.25 AU-0.36 AU, and the actual HZ could be much wider than
this extension, e.g.,  the outer edge of HZ could be 0.36 AU-0.47
AU.  From Fig.4a, if the two giant planets are out of 3:1 MMR,
test particles with initial semi-major axes in HZ are stable,
although their eccentricities will be excited by secular resonance
due to $m_b$ and $m_c$ (Malhotra \& Minton 2008). To see whether
the planet can maintain its orbit in the HZ, we put a planet with
Earth mass in different initial locations. Fig.5 shows the largest
and smallest stellar distances (Q and q ) of the Earth mass
planets. The shaded area is the conservative estimation of the HZ.
 The great variation of $Q$ and $q$ at $a=0.28~\rm AU-0.32~AU$ is
due to the secular resonance of $m_b$ and $m_c$ (Malhotra \&
Minton 2008).
 Interesting to note that, the variation of
 $Q$ and $q$ are small when $\Delta \varpi_{bc}=\pi$, i.e., the
eccentricity excited by the secular resonance of $m_b,m_c$ is
small (Migaszewski et al. 2009). Also, when $m_b,~m_c$ is in 3:1
MMR,  HZ is not in their secular resonance region, so the
variation of $Q$ and $q$ are  also small. No matter in which type
of $\Delta \varpi_{bc}$, most of orbits are in HZ except $a\in
[0.28~ \rm AU, ~0.32~ AU]$ with $\Delta \varpi_{bc}=0,~\pi/2$.
Considering the actual HZ is wider than this, OGLE-06-109L is a
hopeful candidate system for hosting a habitable terrestrial
planet. Simulations in next section indeed show the evidence for
the formation of super-Earth planets in its HZ.

\section{Formation Scenarios of Eccentricity}

In this section, we test the previously mentioned three scenarios
for the eccentricity excitation between the two giant planets in
OGLE-06-109L system: (i) convergent migration and resonance trap
model, where the eccentricities are excited by the trap of 3:1 MMR
during the type II migrations of two planets;  (ii) planetary
scattering model: the eccentricities are generated by close
encounters between some leftover embryos and the planets; (iii)
divergent migration and MMR crossing model, when the
eccentricities are excited by the crossing of either 2:1  or 3:1
MMR during the planetesimal-driven divergent migration. To
simplify the problem, we assume the system has already in its
later stage of formation so that both giant planets have already
formed with their present masses,  coexisting with  tens of
residue embryos with masses in the range of $0.1~ M_\oplus-10~
M_\oplus$ (Earth mass).

For a planet embedded in a geometrically thin and locally
isothermal disk, angular momentum exchanges between the planet and
the gas disk will cause a net momentum lose on the planet,
 which results in a  fast and so called type I migration
  of the planet \cite{GT79,Ward97,Tan02}.
Some mechanisms are proposed recently to reduce the speed or even
reverse the direction of migration. Laughlin et al. (2004),  Nelson \& Papaloizou (2004) proposed that,
in the locations where the magnetorotation instability (MRI) is active,
gravitational torques arising from  megnetohydrodynamical
turbulence will contribute a random walk component to the
migratory evolution of the planets, thus prolong the drift timescale.
%Analytical estimation through Fokker¨CPlanck approach indicates the
%survival rate of planet under type I and stochastic torques is
%around $1-10\%$(Adams \& Bloch 2009).
Paardekooper \& Mellema (2006) noticed that the inclusion of
radiative transfer can cause a strong reduction in the migration
speed. Subsequently investigations (Baruteau \& Masset 2008; Kley
\& Crida 2008; Paardekooper \& Papaloizou 2008) indicated that
 the migration process can be slowed
down or even reversed for sufficiently low mass planets.
Through full 3D hydrodynamical simulations of embedded planets
 in viscous, radiative discs, Kley et al. (2009) confirmed that the migration can be
directed outwards up to planet masses of about 33 $M_\oplus$. Due
to the vagueness  of type I migration, we do not consider this
effect at the present paper.
%Type I migration of these embryos are not considered in
%this work, as Fogg \& Nelson (2007) showed that there will be
%enough solid mass and planetary embryos left under the type I
%migration of embryos and the type II sweeping of  giant planets.
 A detailed study of formation of Earth-like planets in OGLE-06-109L system, which
  includes the type I migration of embryos,  will be
presented in a subsequent paper (Wang \& Zhou, in preparation).

\subsection{Disk model}

According to the conventional core accretion scenario of planet
formation, planet formed through planetesimals coagulation by means
of runaway growth and became protoplanetary embryos by oligarchic
growth in the protoplanetary disk (Safronov  1969; Kokubo \& Ida
1998).
 To model the masses of embryos formed in disk,  we adopt the empirical minimum
mass solar nebula (hereafter MMSN, Hayashi 1981) so that the
surface density of gas disk at stellar distance $a$ is given as
\begin{equation}
\Sigma_{g}=2.4\times 10^3 f _g f_{\rm dep}(\frac{a}{\rm 1AU})^{-3/2}~{\rm g~cm}^{-2}, \label{sigg}
\end{equation}
where $f_g$ is the gas enhancement factor, $f_{\rm
dep}=\exp(-t/\tau_{\rm dep})$ is the gas depletion factor due to
disk accretion, photoevaporation or planet formation with a
timescale of $\tau_{\rm dep}\sim 3$ Myrs (Haisch et al 2001), $t$
is the evolution time.  The surface density of solid disk is given
as,
 \beq \Sigma_{d}=10
f_d \gamma_{\rm ice} (\frac{a}{\rm 1AU})^{-3/2}~{\rm g~cm}^{-2}, \label{sigd}
 \eeq
where $f_d$ is the solid enhancement factor, $\gamma_{\rm ice}$ is
the volatile enhancement with a value of  4.2 or 1 for material
exterior or interior to the snow line (0.68 AU for OGLE-06-109L
system), respectively.
 In such a disk,  the embryos will grow under cohesive collisions
 in a timescale of  (Kokubo \& Ida 2002; Ida \& Lin 2004)
%\beq
\begin{eqnarray}
 \tau_{acc}\simeq 1.6\times 10^5\gamma_{\rm
ice}^{-1}f_d^{-1}f_g^{-2/5}(\frac{a}{\rm 1AU})^{27/10}
\nonumber\\
\times(\frac{M_c}{M_{\oplus}})^{1/3}(\frac{M_*}{M_{\odot}})^{-1/6}
~{\rm yr}, \label{tacc}
\end{eqnarray}

%\end{equation}
where $M_c$ is the core mass. The core growth will continue until it
accretes all the dust material round its feeding zone ($\sim 10$ Hill radii) so
that an isolation body  is achieved with mass of (Ida \& Lin 2004)
\begin{eqnarray}
M_{\rm iso} = 0.12 \gamma_{\rm ice}^{3/2} f_d^{3/2} (\frac{a}{\rm
1AU})^{3/4}(\frac{M_*}{ M_\odot})^{-1/2} M_\oplus. \label{miso}
\end{eqnarray}
For the OGLE-06-109L system, the isolation mass at 4 AU with
$f_d=2$ is around $12~M_\oplus$, above the critical mass
$(\sim 10~ M_\oplus$) for the onset of  efficient  gas accretion
to form  giant planets (Pollack et al. 1996),  and the core growth
timescale is $\sim 1.2$ Myrs.

Interactions between embryos and the gas disk may damp the
eccentricities of the embryos (Goldreich \& Tremaine 1980). The
timescale of the eccentricity-damping for an embryo with mass $m$
can be described as (Cresswell \& Nelson 2006),
\begin{equation}
(\frac{e}{\dot{e}})_{\rm emb}=\frac{Q_e}{0.78}(\frac{M_*}{m})
(\frac{M_*}{a^2\Sigma_g})(\frac{h}{r})^4\Omega^{-1}[1+\frac{1}{4}(e\frac{r}{h})^3]~{\rm
yr }, \label{ede}
\end{equation}
where $r$,  $e$, $h$, $\Omega$ are the stellar distance,
eccentricity of the embryo, scale height of the disk and the
Kepler angular velocity, respectively, $Q_e=0.1$ is a
normalization factor to fit with hydrodynamical simulations.

As an embryo grows to a massive planet ($\ge 30 ~M_\oplus$),  it
will induce strong tidal torques on the disk to open a gap around it
(Lin \& Papaloizou 1993). Then the planet  will be  embedded in the
viscous disk to undergo type II migration. The timescale of type II
migration for a planet with mass $m_p$ can be modelled as (Ida \&
Lin 2004)

\begin{eqnarray}
\tau_{\rm II}=\frac{a}{|\dot{a}|}=0.8 \times 10^6 ~{\rm yr}~
f_g^{-1}(\frac{m_p}{M_{\rm Jup}})(\frac{M_*}{M_{\odot}})^{-1/2}
\nonumber\\
\times(\frac{\alpha}{10^{-4}})^{-1}(\frac{a}{\rm 1AU})^{1/2},
\label{tpii}
\end{eqnarray}

 where $\alpha$ is a dimensionless parameter to adjust the
effective viscosity, and we set $\alpha=10^{-4}$ as a standard value
in our simulation.   When a giant planet is embedded in a gas disk,
tidal interaction of disk may damp its eccentricity if it is not
massive enough. As we mentioned in the abstract,  the situation is
quite elusive for different mass regime  of  the giant planets, so
we adopt an empirical formula (Lee \& Peale 2002) \beq
(\frac{\dot{e}}{e})_{\rm pl}=-{\rm K} \left| \frac {\dot{a}}{a}
\right| \label{edp} \eeq
 to describe the eccentricity-damping rate of giant planets in the OGLE-06-109L system,
 where K is a positive constant with a value ranging $10-100$ \cite{SKK07}.
 After some tests, we choose  $K=10$ in this paper to let
 $e_b$ and $e_c$ have reasonable convergent values.

\subsection{Numerical simulations}

In this section, we simulate the configuration formation for the OGLE-06-109L system with
 N-body models.   We assume the two  giant planets have formed with the observed masses
  at 4 AU and 8 AU-9 AU respectively. The physical epoch corresponds to this assumption
   is  $\sim 1$ Myrs after the formation of star, so that the gas disk is still present.
    The two giant planets have opened  gaps around them  and
     will undergo type II migration according to equation (\ref{tpii})
      in the viscous disk.  At the initial stage of our simulation,
      there are some leftover  embryos in inner orbits that have obtained their
       isolation masses.   To mimic  the formation of Earth-like planets in inner orbits,
       we put 18 embryos,  with masses ranging from $0.17~ M_\oplus$ to $9~ M_\oplus$
        derived from equation (\ref{miso}) and initial
        locations from 0.25 AU to 3 AU. The mutual distances among the embryos are set as
         10 Hill radii. An additional embryo with in situ isolation mass
         will be put between $m_b$ and $m_c$ in model 2.
 All the planets and embryos are initially located
  in near-coplanar and  near-circular orbits ($e=10^{-3}$, inclination $i=e/2$),
   their phase angles (mean motion, longitude of perihelion, longitude of ascending node)
   are randomly chosen.  The acceleration of the planet
    (embryo) with mass $m_i$  is given as,
\begin{eqnarray}
\frac{d}{dt}\textbf{V}_i =
 -\frac{G(M_*+m_i
)}{{r_i}^2}(\frac{\textbf{r}_i}{r_i})~~~~~~~~
\nonumber\\
+\sum _{j\neq i}^N Gm_j [\frac{(\textbf{r}_j-\textbf{r}_i
)}{|\textbf{r}_j-\textbf{r}_i|^3}- \frac{\textbf{r}_j}{r_j^3}]
\nonumber\\
 +\textbf{F}_{\rm edamp}(+\textbf{F}_{\rm migII}),~~~~~~~~~
\label{eqf}
\end{eqnarray}
where $\textbf{r}_i, \textbf{V}_i$ are the position and velocity vectors of $m_i$ in the stellar-centric coordinates,
\begin{equation}
\textbf{F}_{\rm edamp} = -2\frac{(\textbf{v}_i \cdot
\textbf{r}_i)\textbf{r}_i}{r_i^2\tau_e}
\label{fedamp}
\end{equation}
is  damping acceleration,  effective for all embryos and gas
giants but with different $\tau_e$ in equations (\ref{ede}) and
(\ref{edp}), respectively,  and  the acceleration that causes the
type II migration,
\begin{equation}
\textbf{F}_{\rm migII} = -\frac{{\bf V}_i}{2 \tau_{\rm II}}
\label{fmigII}
\end{equation}
is adopted  for two giant planets. We numerically integrate the
evolution of equation (\ref{eqf}) with a time-symmetric Hermit
scheme (Aarseth 2003).  The simulation is performed up to 10 Myrs.
As  we assume that the gas disk depletes exponentially in a
timescale $\tau_{\rm dep}=1$ Myrs, the gas disk almost disappears
at the end of simulation.

During the earlier stage when gas disk is present (the coming
models 1-2), embryos in outer disk will also induce a damping of
giant planets' eccentricities through dynamic friction, which has
similar effect by gas disk. However, as the gas disk dominates
before gas depletion, we did not consider the presence
 of embryos in outer disk in models 1-2, except in model 3 where
 embryos in outer disk are included,  after the gas disk depletes.

%We first perform 28 runs of simulations with 2 planets and 18 (or
%19 in model 2) embryos.  We denote the two giant planets as $m_b$
%and $m_c$ from inner to outer orbits. Table  2 lists the initial
%parameters of three typical  runs corresponding to  models 1 and
%2. The main outcomes  are as follows.

\begin{table*}[htp]
\centering
 \caption{\footnotesize{Initial parameters  for two giant planets
(i-pl) and embryos (i-emb) in  numerical simulations  of model 1
and 2 in section 3.2,  with the outcomes of the survival embryos
(f-emb) and giant planets (f-pl). $P_c/P_b$ is the period ratios
of two giants at 10 Myrs, the end of our simulations.}
\vspace{0.5cm}
 \label{tbl-4}}
\begin{tabular}{lllllll}
\tableline
  ID & i-pl's  & i-emb.'s &  i-emb.'s & i-emb.'s & f-emb's   & f-pl's  a (AU)\\
              &  a (AU) &    No.     &  masses ($M_{\oplus}$)& a (AU) & No. &   and ($P_c/P_b$) \\
\tableline
R1 & 3.8, 8.5  &18  & [0.17, 8.94]          & [0.25, 2.8]       & 4  & 2.53, 5.26; 3:1\\
R2 & 4, 8 & 19  & [0.17, 9.42],  16.8      & [0.26, 3],  6.5     & 1 &  2.40, 5.63; 3.60\\
R3 & 4, 8.2& 19  & [0.17, 9.42],  16.8      & [0.26, 3],  6.5     &11 &  2.83, 5.02; 2.35  \\
%R3 & 4, 8.8 & 20  & [0.17,9.42], 14.9,18.3 & [0.26,3],  5.6,7.3 &4 &  2.46, 6.50; 4.30\\
\tableline
\end{tabular}
%% Any table notes must follow the \end{tabular} command.
%\tablenotetext{a}{Sample footnote for table~\ref{tbl-2} that was
%generated with the \LaTeX\ table environment}
%\tablenotetext{b}{Yet another sample footnote for table~\ref{tbl-2}}
%\tablenotetext{c}{Another sample footnote for table~\ref{tbl-2}}
%\tablecomments{We can also attach a long-ish paragraph of explanatory
%material to a table.}

\end{table*}
{\em Model 1: smooth and convergent migration}. In this model, we
vary the initial locations of two giant planets, $a_b$ and $a_c$.
The 18 embryos with in situ isolation masses in inner orbits are
put so that the outermost one is in an orbit $3.5 $ Hill Radii
away from $m_b$. During the evolution of the typical run R1 ( See
Table 2 for initial parameters),    $m_b$ and $m_c$ are captured
into 3:1 MMR at $t\approx 1$ Myrs, with three resonant angles
liberate  around  either $0$ or $\pi$ with amplitudes around $\sim
0.4~ \pi$ (Fig.6).    Their eccentricities are excited to about
0.1 inside the resonance. Eccentricities of the embryos in the
inner orbits  are also excited due to secular perturbations from
two giant planets, which results in their inward migration in the
gas disk.  At the end of simulation, they merge into 4 planets at
[0.22 AU, ~0.77 AU], with masses   of 4.86 $M_\oplus$, 8.76
$M_\oplus$, 5.79 $M_\oplus$, 16.06 $M_\oplus$ in the order of
increasing semi-major axes.  Noticeably, the inner two are in the
edge of the habitable zone ([0.25 AU, ~0.36 AU]) of the system.

The mechanism that migration of giant planets triggers  the merge
of inner embryos in MMRs  or secular resonances had been already
discussed in many literatures, e.g.,  Zhou et al. (2005), Fogg \&
Nelson (2005). However, in this case,  the eccentricities of
embryos are excited by the secular perturbations of outside giant
planets. The configuration of 3:1 MMR between two  planets are
kept  to the end of the simulation, when they  are stalled at 2.53
AU and 5.26 AU  due to the severe depletion of gas disk. We did 8
runs in this model with $a_b$ varying in [3.5 AU,~ 3.8 AU], and
$a_c$ in [8.2 AU,~ 8.5 AU].  In all the simulations, the  two
giant planets are trapped  in 3:1 MMR, with their eccentricities
osculating around $\sim 0.1$  at the end of evolution as in
Fig.6c.

\begin{figure}[htp]
\vspace{-0.2cm}
 \centering
\includegraphics[scale=0.25,height=2in]{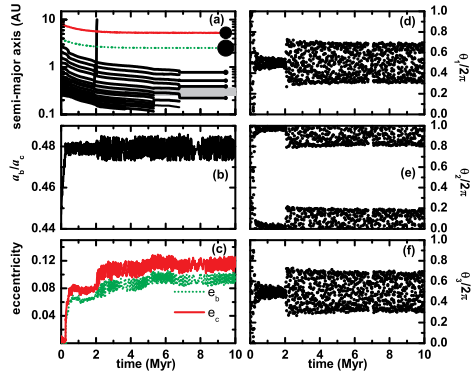}
 \vspace{-0.2cm}
\caption{\footnotesize{Evolution of orbits in model 1 (R1 in Table
2). The two giants ($m_b,~m_c$) are put initially at near-circular
orbits with $a_b=3.8$ AU and $a_c=8.5$ AU.  18 embryos with in
situ isolation masses are put in inner orbits. Panel (a):
Evolution of semi-major axes  of the 2 planets and 18 embryos. The
grey band shows the extension of the habitable zone. The green
dash lines represent the evolution of planet b and the red solid
lines show the result of planet c as the same meaning in panel
(c). Panel (b): Evolution of semi-major axis ratio of two giant
plants. Panel (c): Eccentricity evolution of the two giant
planets. Panel (d, e, f): Evolutions of the resonance angles  for
the 3:1 MMR between $m_b$ and $m_c$.
$\theta_1=3\lambda_c-\lambda_b-2\varpi_b$,
$\theta_2=3\lambda_c-\lambda_b-\varpi_b-\varpi_c$,
$\theta_3=3\lambda_c-\lambda_b-2\varpi_c$. }}
\end{figure}
\begin{figure}[htbp]
\vspace{-0.15cm}
 \centering
\includegraphics[scale=0.25,height=2in]{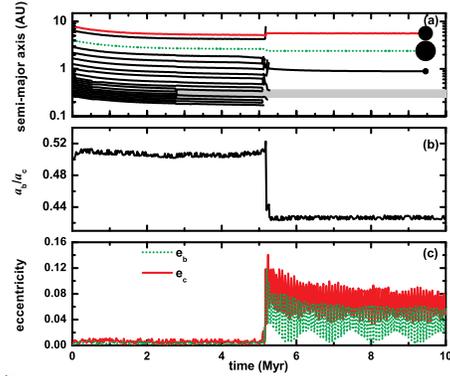}
 \vspace{-0.5cm}
\caption{\footnotesize{Evolution of orbits in model 2 (R2 in Table
2). The two giant planets ($m_b,~m_c$) are put initially at
near-circular orbits with $a_b=4 $ AU and $a_c=8$ AU.  The 18
embryos in inner orbits are put  initially as in Fig.6. An
additional embryo is put at  3.5 Hill radii inside the orbit of
$m_c$. Panel (a): Evolution of semi-major axes  of the 2 planets
and 19 embryos.  The grey band shows the extension of the
habitable zone. The green dash lines represent the evolution of
planet b and the red solid lines show the result of planet c as
the same meaning in panel (c). Panel (b): Evolution of semi-major
axis ratio of two giant plants. Panel (c): Eccentricity evolution
of the two giant planets.}}
 \vspace{-0.5cm}
\end{figure}

{\em Model 2: Planetary scattering during migration}.
     Unlike the
previous model that $m_b$ and $m_c$ undergo smooth migration,  now,
 besides the 18 embryos in inner orbits, we put an
additional embryo with local isolation mass (denoted as $m_{\rm
e1}$, slightly varies for different locations)  between the orbits
of
  $m_b$ and $m_c$.  Fig.7 shows the results of a typical run (R2), with $m_{\rm
e1}=16.8~ M_\oplus$. At  $t \approx 5.2$ Myrs,   a close encounter
between $m_c$ and $m_{\rm e1}$ occurred,  which scatters
 $m_{e1}$ out of the system (Fig.7).
The encounter  excites  $e_c$ up to $0.1$, which in turn excites
$e_b$ from 0 to $\sim 0.08$, resulting in a configuration that
$e_b$  passing $0$ in the eccentricity plane ($e_be_c\cos \Delta
\varpi_{bc}$, $e_b e_c\sin \Delta \varpi_{bc}$) almost
periodically,  the so called near-separatrix (of libration and
circulation of $\Delta \varpi_{bc}$) motion (Barnes \& Greenberg
2006, 2008).
 The eccentricities of embryos in inner orbits
  are also excited due to the sudden increase of $e_b$ and $e_c$,
  which cause  strong mergers
among the embryos into a planet of 14.38 $M_\oplus$ at
0.89 AU.

If  close encounters between $m_{e1}$ and one of the giant planets
occur  much earlier,  tidal interaction between the planets
(embryos) and the  gas disk  will eventually damp their
eccentricities. In run R3,  the scattering process occurred at $t= 1.5$
Myrs when the eccentricity-damping induced by the gas disk is
still strong (Fig.8).  As a result, $e_b$ and  $e_c$ that  excited during planetary scattering
 are damped to less than 0.01 quickly.
   In this case, 11 planets are left with their masses from
0.21  $M_\oplus$ to 16.81 $M_\oplus$  at [0.15 AU, ~1.10 AU].
Among them, 2 embryos are in the habitable zone.

\begin{figure}[htp]
\centering
% \vspace{-2cm}
\includegraphics[scale=0.25,height=2in]{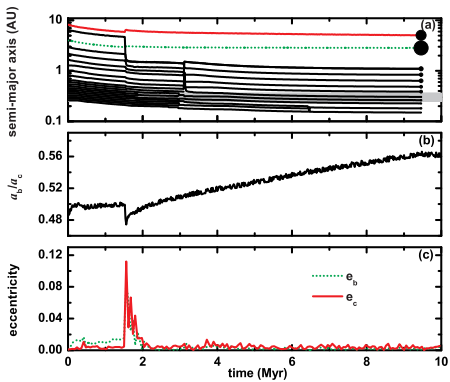}
 \vspace{0cm}
\caption{\footnotesize{Evolution of orbits in model 2 (R3 in Table
2). The two giants ($m_b,m_c$) are put initially at near-circular
orbits with $a_b=4 $ AU and $a_c=8.2 $ AU.  The 18 embryos in
inner orbits are put initially as in Fig.6.
 An additional embryo is put at  3.7 Hill radii inside the orbit
of $m_c$. Panel (a): Evolution of semi-major axes  of the 2
planets and 19 embryos. The grey band shows the extension of the
habitable zone. The green dash lines represent the evolution of
planet b and the red solid lines show the result of planet c as
the same meaning in panel (c). Panel (b): Evolution of semi-major
axis ratio of two giant plants.  Panel (c): Eccentricity evolution
of the two giant planets. }}
\end{figure}

We did 20 runs of simulations with  different  initial locations
of $m_{e1}$.  Among them, 6 runs
  have close encounter events during the evolution, like run R2, resulting in similar
configurations of $m_b$ and $m_c$ as in Fig.7.  5 runs have
earlier close encounters so that $e_b$ and $e_c$
 are damped in the end of simulations. The rest 9 runs  do
not suffer close encounters, but have some  milder encounter
events occasionally. The resulting  $e_b$ and $e_c$ have only slight changes up to $\sim 0.01$.

In a compact configuration with $a_b/a_c > 0.48$,
 semi-major axes  shift to $  a_b/a_c < 0.48$ by planetary scattering is a possible
routine that could lead to the trap  of 3:1 MMR between $m_b$ and
$m_c$. To investigate this  probability, we perform additional 900
runs of simulations with a simplified four-body (star-two
giants-one embryo)  model. In this  model, $m_b$ and $m_c$ are
located initially at 3.7 AU and 7.6 AU so that $(a_b/a_c > 0.48)$
,  with one embryo $m_{e1}$ between their orbits. The initial
semi-major axis of $m_{e1}$  is  chosen at [5.0 AU, ~5.4 AU]. The
mass of $m_{e1}$ slightly varies at different locations, in the
range of $13.8~M_\oplus-14.6~M_\oplus$.
 Both $F_{\rm edamp}$ and $F_{\rm migII}$
 in Eqs. (\ref{fedamp}) and (\ref{fmigII}) are  included.
 Numerical simulations show that, 185 runs out of
900 ones  with close encounter events occurred  between $m_{e1}$
and one of the planets. The epoch for close encounters occurred do
not show a clear  correlation with the relative distance between
the embryo and one of the giant planets (Fig.9).  Among the 185
runs with scattering events,   21 runs ($2.3\%$ of 900 runs)
 lead to  the trap of  $m_b$ and $m_c$ into 3:1 MMR,
 11  runs ($1.2\%$) result in the trap near the boundary of 3:1
MMR,  the rest 153 runs ($17.0\%$) do not  lead to the trap of 3:1 MMR.
  We also observe 62 runs with $m_b$ and $m_c$ being trapped in
8:3 MMR.  Fig.10 shows the final $e_b,~e_c$ at the end of 900
runs' simulations. As we can see, besides those being trapped into
8:3 MMR, $e_b$ and $e_c$ are excited significantly only in those
185 runs with planetary scattering, with the average values $e_b
\sim 0.058$ and $e_c \sim 0.085$ (Fig.10).

\begin{figure}[htp]
\centering
%\vspace{-4cm}
\includegraphics[scale=0.7]{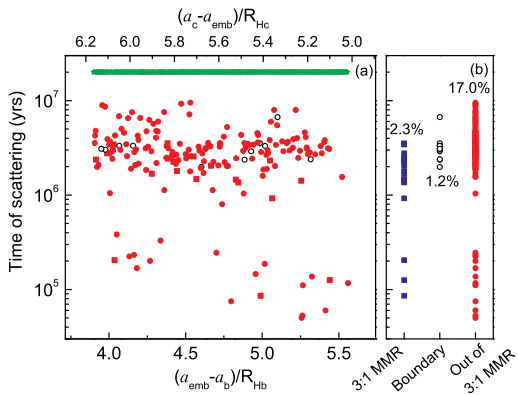}
 \vspace{-0.5cm}

\caption{\footnotesize{Orbital crossing time of the 900
simulations with four-body model (star-two planets-embryo). Panel
(a): Orbital crossing time versus the initial position of the
embryo.
 Separation between the embryo and the planet b measured by the
planet's Hill radius is marked below the figure, while that with
the planet c is marked above the figure. Orbits without close
encounter are represented by green stars; other symbols show the
runs in which close encounters happen, among them blue squares
indicate the runs in which the two planets are trapped into the
3:1 MMR, black circles indicate the cases in which the two planets
are in the boundary of the 3:1 MMR, red circles plot the cases
that the two planets are out of  the 3:1 MMR. Panel (b): the
proportions of each kinds of motions in the total 900 runs of
simulations. }}
\end{figure}

\begin{figure}[htp]
\centering
 \vspace{-1cm}
\includegraphics[scale=0.7]{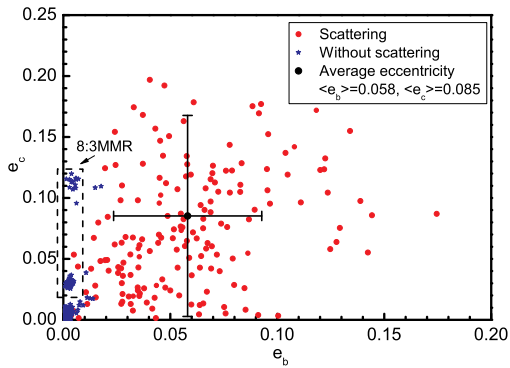}
 \vspace{-0.5cm}
\caption{\footnotesize{Final eccentricities of the two giant
planets in the end of 900 runs of simulations. If no close
encounter happens (indicated by blue stars), the eccentricities of
the two planets are
 small ($\sim$ 0.005) except 62 runs in which the two planets
are trapped into 8:3 MMR. In the 185 runs that close encounters
happen (represented by red circles), eccentricities are excited to
relatively  high values. The averaged  $e_b$ and $e_c$ over the
185 runs with planetary scattering are also plotted, with the
error bars standing for the standard deviations.  }}
\end{figure}

{\em Model 3: Divergent migration in the presence of
planetesimal-disk}. To study the effect of planetesimal-disk in
outside orbits  after the gas disk is depleted, we perform
simulations by including the embryos in outside orbits but
discarding those in inner orbits, since they may be in hot orbits
and have less affections on the outer system. We set two types of
embryos in the outer region, those with masses of
 $5 ~M_\oplus$ and of $0.2 ~M_\oplus$. After some test simulations, we find that the
 solid disk out of 10 AU has little effect on the
 evolutions of the two giant planets at nominal location (See Fig.11 for  a typical run).
So we set the outer edge of solid disk within 10 AU in following
simulations. Two groups of simulations are made according to
different initial $a_b$ and $a_c$.

\begin{figure}[htp]
\centering
% \vspace{-1cm}
\includegraphics[scale=0.35]{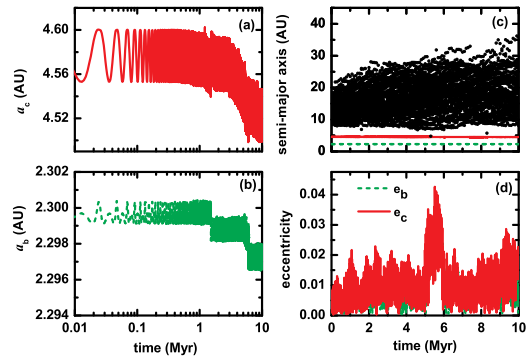}
 \vspace{-0.5cm}
\caption{\footnotesize{Evolution of a typical run in model 3 with
a total disk mass of 27 $M_\oplus$ planetesimal disk outside in
[11 AU, 16 AU]. Panel (a) and (b): evolutions of semi-major axes
of two giant planets. Panel (c): semi-major axis evolutions of all
planets and embryos. Panel (d):  eccentricity evolutions of two
giant planets.
 }}

\end{figure}

{\em Group 3a:} We put initially $m_b$ and $m_c$ at $a_b=3$ AU and  $a_c=4$ AU,
 so $m_c$ is  inside the 2:1 MMR location (at $4.76$ AU) of $m_b$.
 The separation is about $3.3 $ times of their mutual Hill's radii ($R_{Hbc}$),
 above the threshold ($2.4~ R_{Hbc}$), so
 they are Hill stable (Gladman 1993) if there is no other
 perturbations.  We put 63 embryos evenly at [5.5AU,~
9.5AU],  including $5\times 5 ~M_\oplus$  and   $58\times 0.2~
M_\oplus$ ones,
   corresponding to a solid disk of $f_d=2$
  with total  mass of $36.6~M_\oplus$ in [5 AU,~ 10 AU].

\begin{figure}[htp]
\centering
 \vspace{-0.5cm}
\includegraphics[scale=0.25]{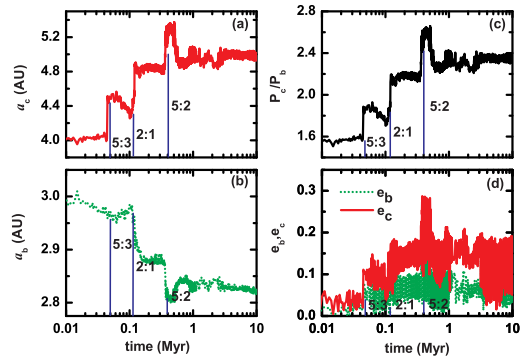}
\vspace{-0.5cm}
\caption{\footnotesize{Evolution of two giant
planets in one run of group 3a (model iii). The blue solid lines
show the epoch of 5:3 MMR, 2:1 MMR and 5:2 MMR orbital crossing
between the two giant planets $m_b$ and $m_c$. Panels (a), (b),
(c), (d) show the evolution of semi-major axes of $m_b$ and $m_c$,
period ratio ($P_c/P_b$), and eccentricities $e_b,e_c$,
respectively. The green dash lines represent the evolution of
planet b and the red solid lines show the result of planet c in
panel (a), (b) and (d).}}

\end{figure}

  Fig.12 shows the evolution of $m_b$ and $m_c$ in a typical run,
  with the innermost embryos $a_{\rm in}=5.5$ AU.
   Under the perturbation of outer embryos, $m_b$ ($m_c$) becomes
   unstable and undergoes inward (outward, resp.) migration quite soon (Fig.12a, b).
    The migration results in
    the quick crossing of $5:3$ and $2:1$ MMR at $t \approx
    0.05$ Myrs, and 0.12 Myrs (see Fig.12c), the two most strong resonances between
  3:2 and 3:1 MMRs, and the system seems to be stable until $m_b$ ($m_c$)
  reaches around 2.82 AU (4.90 AU, resp.), with a drift extension of
  $ 6\%$ ($22\%$, resp.). Their eccentricities are excited after the crossing of
$5:3$ and $2:1$ MMR, with  maximum  values of $\sim 0.2-0.3$
(Fig.12d). Finally $e_b$ and $e_c$ oscillate  in [0.03, 0.15]  and
[0.04, 0.19], respectively. The $5:3$ and $2:1$ MMR crossings of
$m_b$ and $m_c$ lead to the strong scattering of embryos in outer
orbits (Fig.13a), which results in the escape of all the embryos
except one with mass $5~ M_\oplus$ at the orbit of $15$ AU and
eccentricity of 0.45.

  We did 15 runs of simulations in this group, by changing the initial locations of the 63 embryos
  so that $a_{\rm in}$ of the innermost one varies in [5.2 AU,~ 8 AU].
$5:3$ and $2:1$ MMR crossings occurred in 3 runs, with
 eccentricities $e_b \in [0.03,~0.24]$ and $e_c \in [0.04,~0.28]$ at the end of our
 simulations
 ($t$=10 Myrs).
 We did not observe
  resonance-crossings  in  another 3 runs up to
  10 Myrs' evolution, with final eccentricities $e_b \in [0,~0.05]$ and $e_c \in [0,~0.06]$.
  In the rest 9 runs,  $m_b$ and $m_c$ have strong close encounter so that
 $m_c$ is scattered out of the system,
 with the eccentricity of the only survival giant planet $m_b \sim 0.3$.
   The values of $a_{\rm in}$ corresponding to these three types of outcomes (MMR crossing,
   non MMR crossing, $m_c$ being ejected ) do not show
 clear correlation, which indicates the  chaotic states of $m_b$ and $m_c$
 under the perturbation  of outer embryos.

\begin{figure}[htp]
\centering
 \vspace{-0.5cm}
\includegraphics[scale=0.4,width=3in]{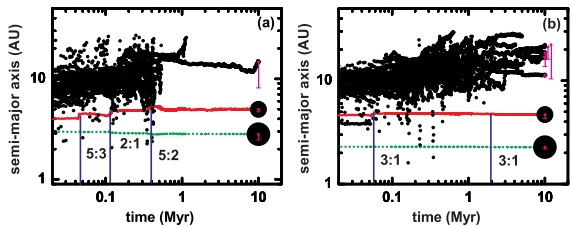}
 \vspace{-2cm}
\caption{\footnotesize{Evolution of two giant planets and embryos
in model 3. The error bar at the end of the evolution show the
extension of $a(1-e)$ of each survival planets and embryos. The
green dash lines represent the evolution of planet b and the red
solid lines show the result of planet c. Panel (a):  semi-major
axis evolutions of one run of group 3a, the blue solid lines
indicate the epoch of 5:3 MMR, 2:1 MMR and 5:2 MMR orbital
crossing between the two giant planets. Panel (b):  semi-major
axis evolutions of one run of group 3b, the blue solid lines
indicate the epoch of 3:1 MMR orbital crossing between the two
giant planets. }}

\end{figure}

\begin{figure}[htp]
\centering
 \vspace{-0.5cm}
\includegraphics[scale=0.25]{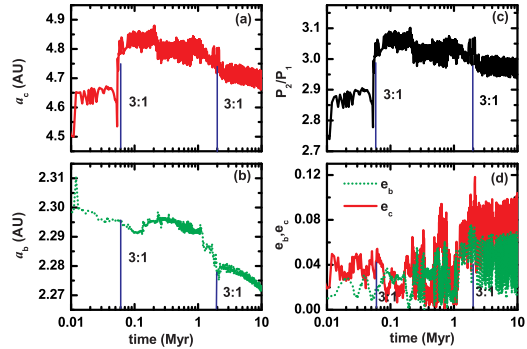}
 \vspace{-0.5cm}
 %\vspace{-5cm}
\caption{\footnotesize{Evolution of two giant planets in one run
of group 3b. The blue solid lines show the epoch of 3:1 MMR
orbital crossing between the two giant planets $m_b$ and $m_c$.
Panels (a), (b), (c), (d) show the evolution of semimajor axes of
$m_b$ and $m_c$, period ratio ($P_c/P_b$), and eccentricities
$e_b,~e_c$, respectively. The green dash lines represent the
evolution of planet b and the red solid lines show the result of
planet c in panel (a), (b) and (d).}} \vspace{-0.5cm}
\end{figure}

{\em Group 3b:} $m_b$, $m_c$ are put initially
 at $a_b=2.3$ AU and  $a_c=4.6$ AU, the nominal locations.
   We put  56 embryos evenly at [6.4 AU, ~10.4 AU],
  including $5\times 5 ~M_\oplus$  and  $ 51\times 0.2 ~M_\oplus$ ones,
   corresponding to a solid disk of $f_d=2$
  with total  mass of $35.2~M_\oplus$ in [5.5 AU,~ 10.5 AU].
 Fig.14 shows the evolution of $m_b$ and $m_c$ in a typical run,
  with the innermost embryos $a_{\rm in}=6.4$ AU.
$m_b$, $m_c$ cross  $3:1$ MMR during the divergent migration
  at $t\approx 0.06$ Myrs and $t\approx 2$ Myrs, causing the increase of $e_b$ and $e_c$,
  At the end of simulation, $m_b$ and $m_c$ locate at $2.27$ AU and $4.71$ AU,
  with $e_b  \in [0.01-0.07]$ and $e_c \in [0.03-0.1]$.
  One $5~M_\oplus$ embryo (at 11.4 AU with $e=0.024$) and $4 \times 0.2~M_\oplus$
  ones (at [16 AU, 22 AU] and e $\in [0.08,~0.53]$) are left (Fig.13b).

   We did  totally 41 runs (13 runs with $f_d=2$ and 28 runs with lower $f_d$) in this group.
    For the 13 runs of $f_d=2$,  5 out of these 8 runs with $a_{\rm in} < 7.8$ AU are
   observed having  $m_b,~m_c$'s  3:1 MMR crossing,
   with final eccentricities $e_b \in [0,~0.08]$, $e_c \in [0,~0.1]$.
 The rest 5 runs with $a_{\rm in} > 7.8$ AU do not have 3:1 MMR crossing,
    with final eccentricities $e_b \in [0,~0.04]$, $e_c \in [0,~0.04]$.
    For the 28 runs with lower solid disks ($f_d=1,~1.2,~1.5,~1.6,~1.8$) and different $a_{\rm
    in}$, no 3:1 MMR crossing is observed in runs with
    $f_d<1.6$,  maybe due to the small
    mass of solid disk ($M<22~M_\oplus$).
     While for $f_d=1.6,~1.8$, the probability of
     3:1 MMR crossing is $\sim 33\%$. The extension of $e_b$ and $e_c$ are similar
     to that of $f_d=2$.

\section{Conclusions and Discussions}

In the paper we investigate the dynamics and formation scenario
for  OGLE-06-109L system in terms of the observed planets
$(m_b,~m_c)$, aiming to understand its formation history and to
predict $e_b$ that is not revealed by observation. According to
the investigation, we find the evolution history of $m_b$ and
$m_c$ depends strongly on the initial conditions, i.e.,
 the cores of $m_b$ and $m_c$
 before  efficient gas-accretion begins.

According to the conventional core-accretion scenario of planet
formation,  a giant planet forms from a massive embryo ($> 10~
M_\oplus$) through accreting nearby gas.  Embryos beyond the snow
line tend to have larger isolated masses, thus they are the ideal
candidates for planetary cores.   However, there will be more than
one embryo beyond the snow line. For example,  assuming  a solid
disk of 2 times of the minimum mass solar nebula (MMSN)  for the
OGLE-06-109L system, the space between 3 AU and  8 AU can be
occupied by  4-5   embryos with isolation masses
 above $10 ~M_\oplus$ and mutual separations $\sim 10$ Hill radii.
Due to the long quasi-hydrostatic sedimentation stage of gas
($\sim $ several Myrs, Pollack et al. 1996),  and the perturbation
from  first generation giant planets to nearby embryos,  all
isolation masses  may have the chance to grow up into second
generation
  giant planets, thus the initial locations of
formed planets can not be well determined.

 For the two giant planets in  OGLE-06-109L system,
 if they formed from embryos with relatively far mutual distances,
 their  initial configuration is  loose, e.g., $a_b/a_c  <  0.48$,
 so that they are beyond the 3:1 MMR. Subsequent
smooth migration under disk tide will  result in a 3:1 MMR,
provided suitable gas depletion timescale. We did 8 runs in model
1,  all the simulations result in the  two giant planets
 trapped in 3:1 MMR, with their eccentricities being excited and
 osculating around $\sim 0.1$ at the end of evolutions as in Fig.6c.

% Their eccentricities   will be excited and remain oscillating around $\sim 0.1$ (Fig.6c).
% However,  close encounters  with the residue embryos  may occur during their
%migration, which will disturb the previous scenario by  shifting
%the semi-major axis ratio,  and thus prevent them into 3:1 MMR in
%most cases. Close encounters with embryos may excite $e_c$ up to
%$0.1$. Subsequent secular evolution between the two planets will
%keep $e_b$ oscillating from around 0 to $\sim 0.06$, the
%near-separatrix motion in the eccentricity plane (Fig.7c).

 If  the two planets formed from embryos with a relatively
small distance,  they may have a compact configuration initially
with  $a_b/a_c  >  0.48$. Then planetary scattering among residue
embryos $(m_{e1})$ and planets are most probably the major cause
of $e_b$ and $e_c$. Among the 900 simulations of model 2 we did,
185 runs ($20.5\%$) with close encounter events occurred between
$m_{e1}$ and one of the planets, with the average values $e_b \sim
0.058$ and $e_c \sim 0.085$ (Fig.10).  Only $3.5\%$ of the 900
runs lead to  the trap (or in the boundary) of  $m_b$ and $m_c$
into 3:1
 MMR.

 After the gas disk is almost depleted, divergent migration of
$m_b$   and $m_c$ caused by the residue embryos and planetesimals  in outer disk may
  drive $m_b$ and $m_c$ passing through lower order MMRs. According to our
  simulations
   of model 3, the crossing of 2:1 MMR is unlikely in the OGLE-06-109L system,
  as it will excite eccentricities of $m_b$ and $m_c$ up to $0.2-0.3$, and it is easy to
  eject $m_c$ out of the system. On the other hand, the crossing of 3:1 MMR is likely,
  which will excite the eccentricities  up to $e_b\sim 0.06$ and $e_c\sim 0.10$. However,  from our simulations, the required
  migration  depends on the mass and radial location of the planetesimal disk.
A solid disk with mass enhancement factor $f_d \ge 2 $ over the
minimum solar nebular may be needed, with their inner edge within
$7.8$AU. Considering the similarities of solar system and the
OGLE-06-109L system, $f_d \sim  2$ is still possible for the
OGLE-06-109L system with a stellar mass $\sim 0.5~ M_\odot$.

In summary, all the three models (i) smooth, convergent migration
and the trap of 3:1 MMR; (ii) planetary scattering; (iii)
divergent migration and the crossing  of 3:1 MMR, $e_b$ and $e_c$
can be excited. However, the probabilities, the conditions and the
final outcomes of these three models are different. Smooth and
convergent migration in model (i), if it occurs as predicted by
the standard model, could result steadily in the trap of 3:1 MMR
between $m_b$ and $m_c$, with $e_b,~e_c\sim 0.1$ all the time. For
model (ii), the probability of planetary scattering occurs is
$\sim 20\%$,  they result in average  $e_b \sim 0.06$ and $e_c
\sim 0.09$, and $m_b$ is more likely to undergo  a near-separatrix
motion in ($e_be_c \cos \Delta \varpi_{bc},~e_be_c\sin
\varpi_{bc}$), i.e., $e_b$ passing 0 at a secular timescale ($\sim
0.08$ Myrs as in Fig.15a). The probability for $m_b,~m_c$ 3:1 MMR
crossing in model (iii)
 depends on the mass and extension of residue solid
disk, and  will result in $e_b \in [0.01-0.07]$ and $e_c \in
[0.03-0.1]$, but the variations in these ranges are in a relative
shorter timescale, e.g.,  $\sim $  0.01 Myrs in Fig.15b.

\begin{figure}[htp]
\centering

\includegraphics[scale=0.25,height=2in]{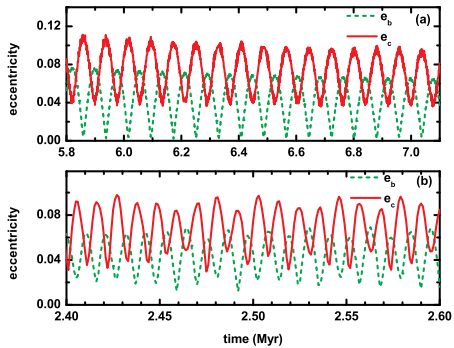}

 %\vspace{-5cm}
\caption{\footnotesize{Eccentricity evolutions of a typical run in
model (ii) and model (iii). Panel (a): zoomed in evolution of
figure 7c from 6 Myrs to 7 Myrs, with  $a_b \approx 2.40$ AU and
$a_c \approx 5.65$ AU.
 Panel (b): zoomed in evolution of figure
14d from 2.4 Myrs to 2.6 Myrs, with  $a_b \approx 2.28$ AU and
$a_c \approx 4.73$ AU. }}

\end{figure}

Some analytical estimations of related timescale is helpful to
reveal the different procedures corresponding to eccentricity
evolution in model (ii) and (iii). The  timescale for the secular
evolution of two planets
  is given as $2\pi/|g_1-g_2|$, where $g_1,~g_2$ are the
 two eigenfrequencies (Murray \& Dermott 1999, Zhou \& Sun 2003).
This gives  0.076 Myrs and 0.037 Myrs for orbits in Fig.15a and
Fig.15b, respectively. In the circular restricted three-body
(CRTB) framework, the timescale of a massless body in the 3:1 MMR
of a perturber is estimated as $2\pi/(ne \sqrt{3\mu' \alpha
f_d(\alpha)})$, where $\mu'=m'/m_*$ is the mass ratio of the
perturber,  $n,e$ is the mean motion and eccentricity of the
massless body, $f_d(\alpha)$ is the function of Laplace
coefficients of semi-major axis ratio $\alpha$ (Murray \& Dermott
1999). Assuming $m_b$ is the perturber,
 $e=0.04$ gives the $e$-evolution timescale  of $m_c$ in $m_b$'s 3:1 MMR as 0.01 Myrs,
 although the CRTB model is not a good model here.
So the eccentricity evolution of Fig.15a is due to the secular
dynamics, while that in Fig.15b is due to the  3:1 MMR crossing.
We also find such a 3:1 MMR timescale is kept for these two orbits
up to the end of simulation.

So to understand scenarios of  eccentricity formation for the
OGLE-06-109L system, we need more detailed information of their
orbits. If $m_b$ and $m_c$ are shown by observation that they are
in 3:1 MMR, then model (i) should be the most possible scenario.
Based on our simulations in section 2.1, we think  the possibility
that $m_b,~m_c$ in 3:1 MMR is small, thus model (i) is unlikely.
However, either model (ii) or (iii) can not be decided by the
present observations. In both models, the averaged  $e_c\sim
0.09-0.10$, roughly agrees with the observed value $
0.11_{-0.04}^{+0.17}$, and  predict the most possible value of
$e_b \approx  0.06$. If $m_b$ and $m_c$ are observed to undergo a
near separatrix motion in ($e_be_c \cos \Delta
\varpi_{bc},~e_be_c\sin \varpi_{bc}$) plane in  a timescale of
secular motion, then
model (ii) is favored for the origin of eccentricities. And we can
infer that,
 after OGLE-06-109L b and c formed, either smooth
migration history during the presence of gas disk is  short,
   or during their migration history, embryos
   may suffer close encounter with the planets, exciting their eccentricities.
 However, if $e_b,~e_c$ oscillates
in  a timescale of nearby (crossed) MMR,
 then most probably  model (iii) accounts for the origin of their
 eccentricities, and based on this  it is possible to predict the extensions of residue disk mass
and location through a more detailed study.

    For the stability of the OGLE-06-109L, the two giant planets
will be stable provided $e_b^2+e_c^2\le 0.3^2$. According to the
formation scenario, super-Earth planets may be formed inside or
outside their orbits.  Numerical simulations show  the region a
$\leq 1.5$ AU (including the habitable zone) or a $\geq 9.7$ AU is
stable. Although the habitable zone contains secular resonance of
the system (Migaszewski et al. 2009; Malhotra \& Minton 2008), our
investigation shows that it is wide enough for an Earth-mass
planet being formed and stable at least 10 Myrs. In the rare cases
when the two giant planets are in the 3:1 MMR, the stable region
in  inner orbits  is reduced to $a<1.4$ AU, while that in the
outer region is enlarged to $a<7.5$ AU.

  When extending the analyses  to other multiple-planet systems,
   as close encounters between residue embryos are common, we
expect planetary scattering and the consequent near-separatrix motion
 of eccentricities among multiple planetary systems are
also common, which agrees with the statistics of the multiple
exoplanet systems observed (Barnes \& Greenberg 2006, 2008).
 Divergent migration is also possible to sculpt the architecture of the
multiple planetary systems. Both  mechanisms can account for the
presence of modest eccentricities
 in the multiple planetary systems without necessarily being trapped in MMRs.

We thank Dr. S. Mao for useful discussions, and the anonymous
referee for his constructive suggestions. This work is supported
by NSFC (10925313, 10833001, 10778603), National Basic Research
Program of China (2007CB814800).

%\end{multicols}
% [Adams \& Bloch 2009]Adams, F. C., Bloch, A. M.,ApJ,  701, 1381
%Muto, T., Inutsuka, S., 2009,  ApJ 701, 18
%Paardekooper,S.-J.,  \& Nelson, R.P, 2009, Invited Review to
%appear in Advanced Science Letters (ASL),
% Special Issue on Computational Astrophysics, edited by Lucio Mayer, eprint arXiv:0906.4347
\clearpage

\end{document}